\documentclass[12pt,preprint]{aastex}

\usepackage{natbib,aas_macros}

\shorttitle{A ROSAT BSC Survey with Swift}
\shortauthors{Fox}

\newcommand{\swiftbscs}{\mbox{Swift-BCS}}

\newcommand{\yesxmm}[2]{}

\newcommand{\chandra}{\textit{Chandra}}

\newcommand{\einstein}{\textit{Einstein}}
\newcommand{\rosat}{\textit{ROSAT}}

\newcommand{\xmm}{\textit{XMM}}
\newcommand{\xmmlong}{\textit{XMM-Newton}}

\newcommand{\swift}{\textit{Swift}}

\newcommand{\ctsec}{\mbox{c s$^{-1}$}}
\newcommand{\cmsq}{\mbox{cm$^{2}$}}

\newcommand{\ergcms}{\mbox{erg cm s$^{-1}$}}

\newcommand{\usno}{USNO-A2}

\newcommand{\lxlopt}{\mbox{$L_{\rm X}/L_{\rm opt}$}}

\newcommand{\tee}[2]{\mbox{$#1 \times 10^{#2}$}}

\def\simlt{\mathrel{\hbox{\rlap{\hbox{\lower4pt\hbox{$\sim$}}}\hbox{$<$}}}}
\def\simgt{\mathrel{\hbox{\rlap{\hbox{\lower4pt\hbox{$\sim$}}}\hbox{$>$}}}}

\def\ale{\mathrel{\hbox{\rlap{\hbox{\lower4pt\hbox{$\sim$}}}\hbox{$<$}}}}
\def\age{\mathrel{\hbox{\rlap{\hbox{\lower4pt\hbox{$\sim$}}}\hbox{$>$}}}}

\begin{document}

\title{A \rosat\ Bright Source Catalog Survey with the \textit{Swift}
  Satellite} 

\author{Derek B. Fox}
\affil{Caltech Astronomy, MS 105-24, 
       California Institute of Technology, Pasadena, CA, 91125-2400;
       \textit{derekfox@astro.caltech.edu}}


\begin{abstract}

We consider the prospects for a complete survey of the 18,811 sources
of the \rosat\ All-Sky Survey Bright Source Catalog (BSC) with NASA's
\swift\ gamma-ray burst (GRB) mission.  By observing each BSC source
for 500\,s with the satellite's imaging X-ray and UV/optical
telescopes, this ``Swift Bright (Source) Catalog Survey'' (\swiftbscs)
would derive $<$5\arcsec\ source positions for all BSC sources that
have not faded substantially from their \rosat-era flux levels, and
provide 0.2--10 keV X-ray and UV or optical flux measurements.  The
improvement by a factor of 10 to 30 in the two-dimensional
localization for these sources will enable optical identifications or
deep limits in nearly every case, fulfilling the promise of the BSC as
a multiwavelength catalog, and allowing the full enumeration of its
rare source populations.  Since the \swiftbscs\ can be accomplished
with 10\% of the time on-orbit, for a three-year mission, and since
its targets will be of lower priority than active afterglows, it will
not conflict with the primary \swift\ mission of GRB follow-up
observations.  Moreover, the BSC targets can be scheduled so as to
advance the secondary goals of \swift: Daily monitoring of the full
sky in the hard X-ray band (source fluxes of $\simgt$20\,mCrab,
10--100\,keV) with the wide-field Burst Alert Telescope (BAT); and a
two-year all-sky BAT survey down to $\simgt$1\,mCrab.  The resulting
expansion of the catalog of identified X-ray sources from 2000 to
18,000 will provide a greatly-enriched set of targets for observation
by \xmmlong, \chandra, and future high-energy observatories.

\end{abstract}



\section{Introduction}
\label{sec:intro}

NASA's \swift\ Midex mission is in its final ground-based testing
phases and is scheduled for launch in September 2004\footnote{\swift\
mission web site: \texttt{http://swift.gsfc.nasa.gov/}}.  Designed for
and dedicated to the detection and automated follow-up of gamma-ray
bursts (GRBs), \swift\ has a broad-band complement of instruments: a
coded-mask CdZnTe Burst Alert Telescope (BAT; \citealt{barthelmy00})
to localize the prompt hard X-ray emission of gamma-ray bursts, and
for pointed observations of the subsequent GRB afterglows, a
nested-mirror grazing-incidence X-ray telescope (XRT;
\citealt{bhn+03}) and a UV/Optical telescope (UVOT; \citealt{rtn+00})
which is a slightly-modified version of the Optical Monitor on
\xmmlong\ \citep{mbm+01}.  The XRT, with an EPIC MOS CCD as its focal
plane detector, has an effective area of 135 \cmsq\ at 1.5\,keV,
making it the seventh-largest focusing X-ray telescope ever flown --
including the three telescopes on \xmm\ -- and its 18\arcsec\
(half-power diameter) point-spread function compares favorably with
all previous missions except the \einstein\ and \rosat\ HRI
instruments, \chandra, and \xmm.  The UVOT provides UV/optical
(200\,nm to 600\,nm, depending on the filter) sensitivity over most of
the field-of-view of the XRT, and can reach $B = 24$~mag in a 1000\,s
exposure.

Uniquely among missions of its size, \swift\ possesses large reaction
wheels which give it a slew rate of one degree per second, enabling
prompt follow-up of GRB positions and a continuous sequence of
pointings over the course of each orbit, scheduled so that the
narrow-field instruments (XRT and UVOT) are never Earth-occulted.  In
combination, as we explore below, these capabilities make it
attractive and feasible to execute a complete \swift\ survey of the
\rosat\ Bright Source Catalog.

The \rosat\ All-Sky Survey (RASS) Bright Source Catalog (BSC;
\citealt{vab+99}) contains 18,811 sources, covering 92\% of the sky,
down to count rates of 0.05 PSPC \ctsec, corresponding to a flux limit
of roughly \tee{7}{-14} \ergcms\ (0.5--2.0 keV) for an unabsorbed
blackbody spectrum with $kT=0.1$~keV.  It is the most sensitive X-ray
all-sky survey ever performed, reaching fluxes 100~times fainter than
the \einstein\ survey, and with a greater emphasis on the soft X-ray
band.  Typical flux-dependent positional uncertainties for BSC sources
are 10--20\arcsec\ (radius, one-sigma).

These positional uncertainties are large enough that they have
prevented confident optical identification -- and hence, physical
classification -- of the great majority of BSC sources.  Firm
identifications have been derived for a subset of roughly 2000 of the
brightest sources by the \rosat\ Bright Survey group
\citep{fhs+98,shl+00}. Suggested identifications have been proposed
for perhaps 4000 more in wide-area optical surveys
\citep{zeh+03,avm+03}, but the confidence level of these is such that
most have not been subject to subsequent follow-up observations.

Among the $\simgt$12,000 unidentified BSC sources that remain there
must exist, undiscovered and in quantity, members of rare and
interesting X-ray source populations; members of common source
populations that exhibit distinctive features making them worthy of
further study; and quite possibly, entirely new classes of Galactic or
extragalactic X-ray source that have so far escaped notice.

These rare and interesting source populations are best identified from
within the BSC rather than by other methods, for at least two reasons.
First, any BSC source has a sufficiently high X-ray flux to permit
detailed follow-up observations by \xmm, \chandra, and future
high-energy missions.  Second, as we discuss below, the optical
counterparts of BSC sources are bright enough, in most cases, to be
accessible to optical spectroscopy from modest-sized (\mbox{4-m}
class) facilities, which will allow extensive ground-based exploration
of the physical properties of the sources.

In the sections that follow, we make the case for a complete survey of
BSC sources with \swift, the \swiftbscs.  With a \mbox{500-s} pointing
on each source, sources near the brightness threshold of the BSC will
be re-detected with 25 counts in the \swift\ XRT, yielding a
$<$5\arcsec\ position from the X-rays alone.  The simultaneous 500-s
UVOT exposure will have a detection limit of $B=23.5$ in the
$B$-filter, with comparable limits in other UV/optical filters when
they are used.  In most cases this combination of observations will
enable the identification of a unique UV/optical counterpart for the
source, allowing flux estimates or upper limits for the source to be
derived from archival surveys spanning the electromagnetic spectrum.
In all, the required 9.4~Msec of observations can be accomplished with
10\% of the time on-orbit over a three-year portion of the \swift\
mission.  It will thus be possible to schedule \swiftbscs\
observations so as to advance the secondary goals of the mission:
Daily monitoring of the full sky to fluxes of $\simgt$20\,mCrab
(10--100\,keV) with the BAT, and a two-year all-sky BAT survey to
$\simgt$1\,mCrab.


\section{Fulfilling the Promise of the BSC}
\label{sec:promise}

In this section we show how deriving a uniform set of $<$5\arcsec\
precision source localizations for the sources of the BSC will fulfill
the promise of the BSC as a multiwavelength catalog of the 18,811
brightest soft X-ray sources in the sky.  In addition, we discuss past
and current efforts at BSC source identification, and outline some of
the returns that near-complete BSC identification will enable.


\subsection{BSC Optical Counterparts}
\label{sub:promise:optical}

The identification of optical counterparts for every BSC source, or
the establishment of constraining upper limits, is the crucial
enabling step that will convert the BSC from an X-ray source catalog
into a multiwavelength catalog which -- by comparison to large-area
off-band surveys -- can present the X-ray, optical, near-infrared, and
(in many cases) radio properties of those same X-ray sources.

Most known types of X-ray sources in the BSC will have optical
counterparts that are bright enough to have been detected in the
Palomar All-Sky Surveys and to be listed in the resulting USNO%
\footnote{USNO star catalogs web site:
  \texttt{http://ftp.nofs.navy.mil/projects/pmm/catalogs.html}}
and DPOSS%
\footnote{DPOSS web site: \texttt{http://dposs.caltech.edu/}}
catalogs.  Such sources will also be accessible to
optical-spectroscopic investigation from modest-sized ground-based
facilities.  At the limiting BSC count rate of 0.05~PSPC \ctsec, the
expected source $B$-band magnitude may be expressed approximately as a
function of \lxlopt:
\begin{equation}
    B \approx 17.5 + 2.5 \log_{10} (\lxlopt) \mbox{ mag.}
\end{equation}
Typical values of $\log_{10}(\lxlopt)$ for various source populations
are: for stars, $-1$ or less; for galaxies, $-0.25$ or less; for AGN
and quasars, $-1$ to $+1$; and for white dwarfs and X-ray binaries,
$+1$ to $+3$ (in extreme cases).  At the limit defined by the X-ray
binaries, $B\approx 25$~mag, the counterparts will not be present in
the POSS images, and ground-based spectroscopy will require \mbox{8-m}
class telescopes, but these cases are expected to be relatively rare.
Moreover, within regions covered to the deeper limits of the Sloan
Digital Sky Survey (SDSS; \citealt{yaa+00}) even X-ray binary
counterparts may be detected in reasonable numbers.

The only known optically-inaccessible source population consists of
the isolated neutron stars (INSs), which have $\lxlopt\sim 10^5$
\citep{wm97,ttz+00,krh04} and are therefore too faint for optical
spectroscopy with any current facilities.  However, these sources are
interesting in their own right: the INSs have been subject to
intensive observation -- including more than a megasecond of
high-spectral resolution observations with \chandra\ and \xmm\ -- as
they are promising test-beds for theories of neutron star (NS)
atmospheres, NS structure, the properties of bulk matter at
super-nuclear densities, strong-field gravity, and (depending on their
intrinsic magnetic field strengths) strong-magnetic field quantum
electrodynamical effects.  

\begin{figure}[t]
\plotone{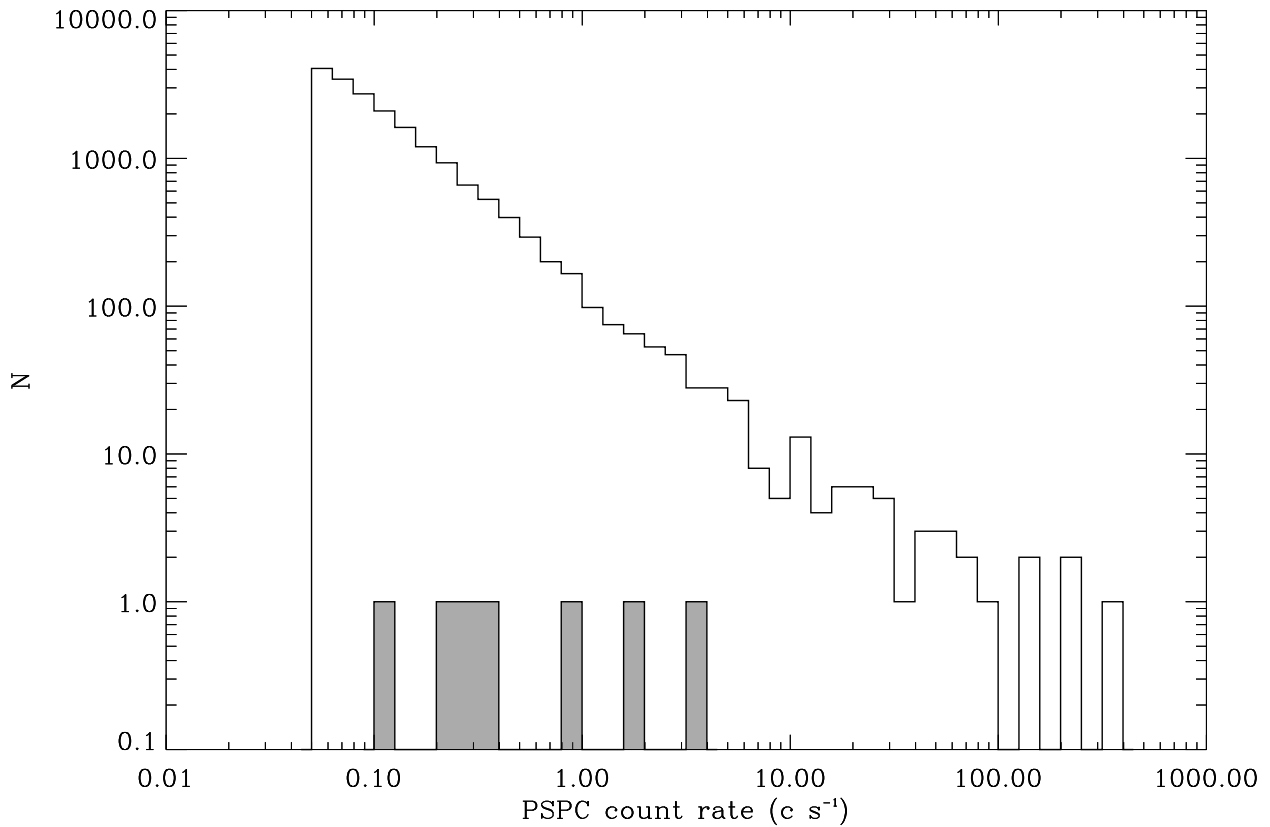}
\caption[]{\small%
\rosat\ BSC source luminosity function, and comparison to the known
population of isolated neutron stars (shaded histogram;
\protect\citealt{ttz+00,krh04}).  With the vast majority of BSC sources
unidentified, many INS sources may be present within the fainter
portions of the BSC.  Note that the power-law slope of the luminosity
function at low count rates is $\alpha = -1.23$.
}
\label{fig:bsc-hist}
\end{figure}

Given the relatively bright X-ray fluxes of the seven INSs that are
known, many more should be present in the BSC down to the survey
threshold (Fig.~\ref{fig:bsc-hist}).  But the relatively large BSC
positional uncertainties mean that optically-bright objects may be
found with high probability in every error circle: \usno, down to
$B\sim19$, has roughly 10 optical objects in each typical
60\arcsec-diameter (2-sigma) region.  Candidate INSs, which should be
apparent as BSC sources without likely optical counterparts, are
therefore hard to identify.  

This same confusion problem has prevented the routine identification
of BSC optical counterparts even though -- as we have seen -- most of
those counterparts already exist in current optical survey catalogs.
Refining the BSC positions to $<$5\arcsec\ precision, however -- a
factor of $\approx$3 improvement over the BSC -- will provide a factor
of $\approx$10 improvement in the two-dimensional localization,
reducing the average number of \usno\ objects per 2-sigma localization
from 10 to one, and thus enabling unique optical identification in
nearly every case.

In short, for each $<$5\arcsec\ BSC source position, either a
high-likelihood optical counterpart will be apparent, or the source
will be an immediate candidate high-\lxlopt\ object, that is, a likely
X-ray binary or INS.

The systematic identification of the BSC optical counterparts will, in
turn, realize the potential of the BSC as a multiwavelength catalog.
The \usno\ and DPOSS will provide $B\!RI$ color information for the
sources, and the 2MASS catalog%
\footnote{2MASS web site:
  \texttt{http://www.ipac.caltech.edu/2mass/}}
will provide $JHK_s$ color information.  Comparison to the large-area
NVSS and FIRST radio surveys will allow identification of radio
sources.  And finally, the identification of the optical counterparts
will enable ground-based spectroscopic studies of the sources, most of
which will be accessible to \mbox{4-m} class facilities. 


\subsection{Past and Present Efforts}
\label{sub:promise:past}

BSC follow-up efforts to-date, although extensive, have ultimately
yielded firm identifications for only a minority of the BSC sources.
The Hamburg/RASS Catalog (HRC; \citealt{zeh+03}) contains suggested
identifications for 4388 BSC sources based on examination of digitized
direct and objective-prism Schmidt plates covering approximately
10,000\,deg$^2$ of the Northern sky, excluding regions within 30\degr\
of the Galactic plane.  This represents an identification rate of 82\%
for the 5341 BSC sources in the region; however, the contamination
rate of the catalog -- the number of proposed counterparts which are
actually misidentifications -- is not well known.  \citet{bev+98} have
estimated a contamination rate of 2\% for a preliminary version of the
survey; however, a quantitative basis for this estimate is not given,
and it seems likely to be an underestimate.  Indeed, since the catalog
method is to identify each BSC source with its most plausible optical
counterpart within 30\arcsec\ -- based on a priori expectations of the
properties of various X-ray emitting source populations -- the
identification of new or unusual source types is nearly excluded by
definition.

The \rosat\ Bright Survey (RBS; \citealt{fhs+98,shl+00}) has pursued
optical identifications for the bright portions of the BSC ($>
0.2$\,\ctsec), also excluding regions within 30\degr\ of the Galactic
plane; to date they have achieved a 99.5\% identification rate for
approximately 2000 BSC sources.  Again, the contamination fraction is
not known.  Two known INS sources, RX~J1308.6+2127 and
RX~J1605.3+3249, were discovered in the RBS as BSC sources without
bright optical counterparts \citep{shs+99,mhz+99}.

As part of the ongoing Sloan Digital Sky Survey, a component RASS/SDSS
survey \citep{avm+03} will be taking spectra of roughly 10,000
candidate counterparts to RASS sources drawn from the BSC and its
companion Faint Source Catalog (FSC; \citealt{vab+00}).  Sources in
the Sloan survey region, with detection likelihoods of more than 10 in
either the BSC or FSC, will be targeted with (on average) one
spectroscopic fiber per source, placed on a single candidate optical
counterpart brighter than the SDSS spectroscopic limit ($g$, $r$, or
$i<20.5$~mag).  This approach, combined with standard SDSS 5-color
quasar selections, has proved efficient at identifying likely X-ray
luminous AGN \citep{avm+03}; at the same time, it cannot address the
properties of the optically fainter populations in the BSC.

Finally, during the lifetime of the \rosat\ mission itself, a number
of HRI surveys of BSC targets, selected by various metrics, were
carried out.  The aim of these surveys was typically to improve the
source positions, and/or resolve the sources themselves, with
higher-resolution HRI imaging; since the HRI is less sensitive than
the PSPC, pointed observations of at least 1\,ks exposure time are
required for these purposes.  In all, such HRI pointings provide
coverage for 9.1\% of the BSC sources, and the varying motivations for
the surveys makes this subset of BSC source identifications a somewhat
motley one.

\citet{rfb+03} investigated the prospects for BSC source
identification using statistical techniques of catalog
cross-correlation (see also \citealt{rbp+00}).  In particular, they
compared the BSC to the \usno, IRAS point-source, and NVSS catalogs,
and carried out follow-up observations of candidate high-$L_{\rm
X}/L_{\rm opt/IR/radio}$ sources with the \chandra\ HRC and
ground-based telescopes, with the aim of either discovering new INSs
or setting stringent limits on their frequency within the BSC.
Although they find no new INSs, they ``rediscover'' two: with a high
probability of having no counterpart in the off-band catalogs, these
two are selected as part of the authors' original 32-candidate sample.
Moreover, by applying their selection to a group of artificial test
sources, they show that their sample is subject to an 80\% confusion
rate: 80\% of isotropically-distributed sources without off-band
counterparts are nonetheless associated with a nearby, bright optical
object with modest ($<$90\%) confidence.  Naturally, this is most
commonly true of sources in the Galactic plane, where the density of
\usno\ objects is highest.

Perhaps more importantly, \citet{rfb+03} find that all 12 of the X-ray
sources for which they are able to determine refined positions, either
from \chandra\ HRC (sub-arcsec) or \rosat\ HRI ($\sim$3 arcsec)
observations -- and none of which are INSs -- are associated with
objects from the \usno\ optical and/or 2MASS near-infrared (NIR)
catalogs.  Thus, they conclude that the identification of off-band
counterparts to BSC sources can be seen as limited strictly by the
absence of higher-quality X-ray positions, rather than by high X-ray
to optical/NIR flux ratios.  This confirms the argument we have made
by reasoning from the \lxlopt\ properties of known source populations:
given the X-ray flux limit of the BSC, the existence of off-band
counterparts in archival surveys can be expected, for all known source
populations except the X-ray binaries and INSs.


\subsection{Complete BSC Identification}
\label{sub:promise:complete}

Identifications of the BSC sources, as a class, are worth pursuing.
The very brightness of these brightest soft X-ray sources makes them
interesting, since they can easily be made the subject of high
signal-to-noise follow-up observations with \chandra, \xmm, and future
high-energy missions such as \textit{Constellation-X}.  In particular,
any object at the BSC threshold (depending on the source spectrum)
will yield 3000 to 6000 or 1500 to 3000 counts in the dispersed
spectrum of a 100-ks exposure with the \xmm\ or \chandra\ gratings,
respectively, and count rates of $\approx$0.7~\ctsec\ (\xmm\ EPIC PN)
or $\approx$0.15~\ctsec\ (\chandra\ ACIS-S) in the main CCD detectors
of those missions.  

The large-scale efforts that have already been devoted to optical
follow-up and identification of BSC sources, discussed above, testify
to the intense interest in establishing optical counterparts to these
sources.  Improved positions for BSC sources will greatly assist these
efforts, providing the impetus that is probably necessary to drive
them forward to an ultimate completion level of $\simgt$90\%.  Efforts
to-date have focused primarily on the bright end of the BSC and away
from the Galactic plane, and with good reason -- at fainter flux
levels and nearer the plane confusion problems for the BSC positions
are severe.  Only improved positions from X-ray observations can
alleviate this difficulty.

Achieving a substantial completion fraction for BSC source
identifications is the best means to enumerate in full the rare source
populations of the BSC and identify new source populations if they
exist.  The large-area surveys of the BSC have been biased towards
discovery of optically bright or emission-line sources associated with
X-ray bright BSC sources outside of the plane of the Galaxy
($|b|>30\degr$).  This focus on high Galactic latitudes presents a
particular problem for Galactic populations, which can be expected to
prefer the plane of the Galaxy: such populations will be subject to
greater confusion in general and, in addition, will have had a large
fraction of their population -- all those members within 30\degr\ of
the plane -- skipped entirely by most follow-up efforts.  Apart from
the INSs, such populations could include, for example, low-mass X-ray
binary systems (LMXBs) in quiescence, and Bondi-accreting isolated
black holes, the higher-mass counterparts to the INS population.

The discovery of sources in new or rare extragalactic populations is
also a possibility.  For the most part, the BSC surveys have operated
under the presumption that any AGN within a BSC error circle is
necessarily a BSC source counterpart.  While the estimated sky density
of AGN is low enough to make this approach reasonable in the
aggregate, without any quantitative estimate of the likelihood of each
association, contamination of the resulting sample is guaranteed.  In
each case where such a mistaken BSC-AGN association has been made, we
will find instead a non-X-ray luminous AGN, on the one-hand, and a BSC
source without prominent AGN counterpart, on the other.  The latter
class of BSC sources may well yield new types of extragalactic X-ray
sources.  

Finally, as we discuss further in the sections that follow, there are
good reasons to perform BSC follow-up with \swift.  In addition to
providing improved positions for the BSC sources, a \mbox{500-s}
observation with the XRT and UVOT will provide new spectral
information in the 0.2--10~keV X-ray band and in the optical or UV
(200 to 600 nm), depending on the choice of UVOT filter.  Since the
BSC sources are prominent in the soft X-ray band, they are an
exclusively low-extinction population, and the UV data -- which cannot
be gathered by ground-based observers -- are of particular interest.
Whenever possible, one of the UV filters should be in place for BSC
observations; it is not clear if this will be possible, however, in
every case (\S\ref{sub:swiftops:constraints}).


\section{Observational Strategy}
\label{sec:strategy}

Our proposed observational strategy is straightforward.  Since the
\swift\ XRT and \rosat\ PSPC have comparable sensitivity and
overlapping spectral ranges (Table~\ref{tab:props}), a broad range of
spectra produce comparable count rates in the two instruments.  In
particular, for a given PSPC count rate, very soft spectra produce
fewer counts in the XRT, and very hard or extinguished spectra produce
many more counts in the XRT.

We therefore choose a standard \mbox{500-s} exposure per source.  This
will suffice to collect $\sim$25~counts from sources near the
0.05~\ctsec\ threshold of the BSC which have not faded substantially
from their \rosat-era flux level.  Detection of 25~counts will provide
a detection with high confidence, and allow a localization with a
purely statistical precision of 1.5-arcsec (radius, 1-sigma).  It is
anticipated that the absolute pointing of the XRT, relative to the
UVOT, will be subject to systematic uncertainties at the 3-arcsec
level; however, it is possible that experience with satellite
operations on-orbit will eventually reduce this systematic uncertainty
to less than 1~arcsec.

\begin{figure}[t]
\epsscale{0.9}
\plotone{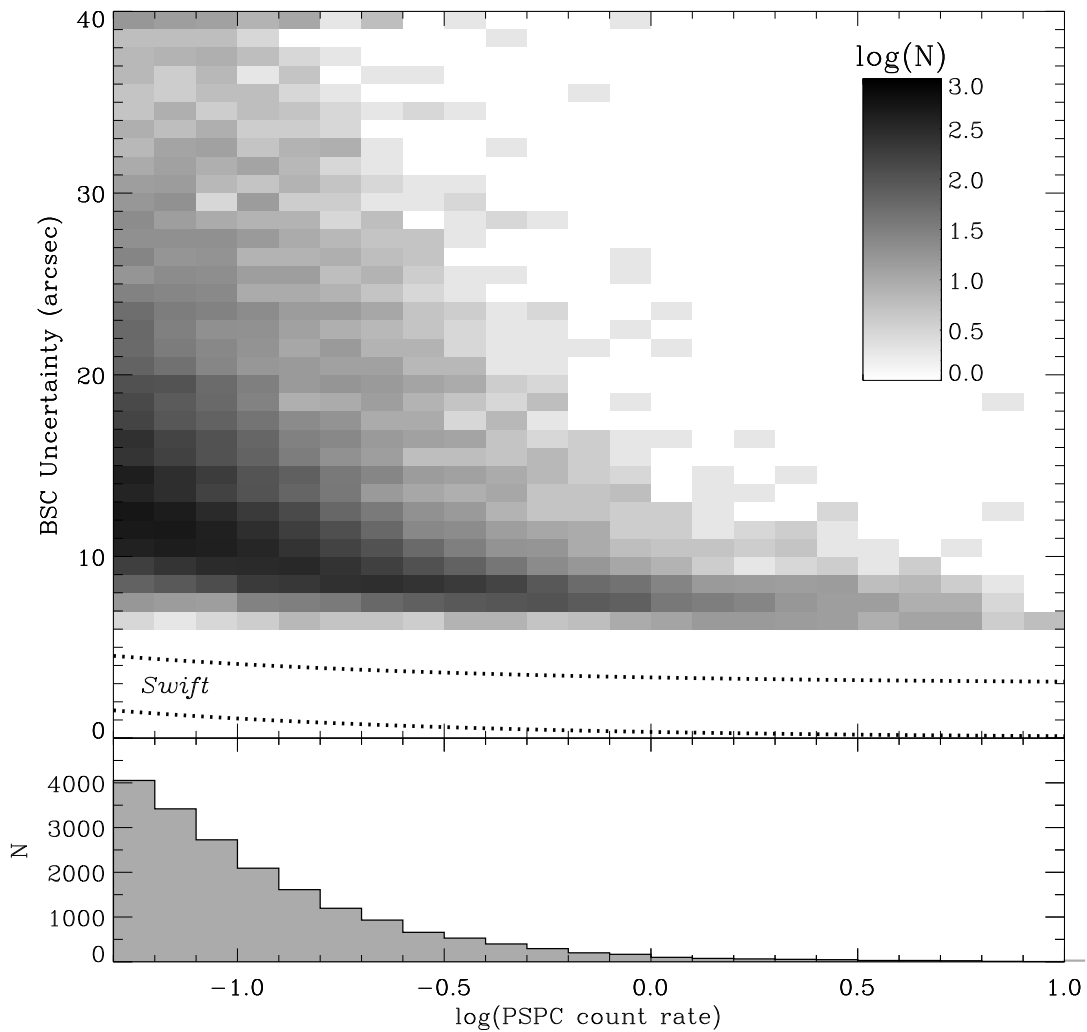}
\caption[]{\small%
Distribution of \rosat\ BSC sources in count rate and positional
uncertainty.  Top panel: Two-dimensional histogram of BSC sources as a
function of count rate and positional uncertainty, represented in
grayscale (note the logarithmic stretch).  Approximate \swift\ XRT
uncertainties (500\,s exposure) as a function of source count rate are
indicated by the dotted lines; the lower line indicates the
statistical uncertainty only, with the upper line including an
estimated 3-arcsec systematic uncertainty.  Most classes of source are
expected to yield roughly equal count rates in the XRT and PSPC.
Bottom panel: One-dimensional histogram of BSC sources as a function
of count rate.  For the great majority of BSC sources, a \swift\ XRT
snapshot will improve the position determination by a factor of 3 to 4
in diameter and 10 to 15 in area.
}
\label{fig:bsc-gray}
\end{figure}

Figure~\ref{fig:bsc-gray} illustrates the two-dimensional distribution
of BSC sources in PSPC count rate and localization uncertainty
(radius, 1-sigma).  Under the assumption of count-rate parity with the
\swift\ XRT, it also shows the expected positional uncertainties from
our standard \mbox{500-s} exposure, both with and without the 3-arcsec
contribution from the XRT absolute pointing uncertainty.  As the
figure demonstrates, BSC sources that have more or less maintained
their \rosat-era X-ray luminosity will be localized with 10 to 30
times better two-dimensional precision than in the BSC.

These uncertainty calculations will be rendered academic in cases
where a counterpart can be identified (and hence localized to
sub-arcsec precision) in the simultaneous UVOT exposure, which will
have a limiting magnitude of $B=23.5$~mag for the $B$ band, and a
comparable limit for other filters when they are used.  Since BSC
source positions are already subject to confusion at the limit of the
\usno\ catalog and corresponding archival optical surveys, the
identification of a high-likelihood counterpart from UVOT imaging
alone will be most probable in cases where a UV or far-UV filter is
used, since this will help to distinguish the BSC sources from the
field.

In other cases, a UVOT identification may follow from the refined
localization derived from the XRT data.  The potential for this type
of ``localization cascade,'' deriving from a single pointed
observation with a single satellite mission, is a unique but not
accidental capability of \swift, which was designed to produce this
cascade routinely in the course of its rapid-response GRB follow-up
observations.

Survey observations are not time-sensitive, so given the rapid one
degree per second slew rate of the \swift\ spacecraft, the scheduling
of the \swiftbscs\ should not present significant operational
challenges.  For example, if we demand a duty cycle of $\simgt$90\%
for survey operations, then we are restricted to slewing, on average,
less than 25\degr\ before and after each BSC target.  This should not
be a difficult requirement to meet; smaller average slews can probably
be achieved with optimization of the planning software.


\section{\swift\ Operations and the Swift-BSCS}
\label{sec:swiftops}

The current plan for \swift\ operations on-orbit is summarized in the
Technical Appendix to the announcement and call for proposals for the
\swift\ Cycle 1 Guest Investigator Program.%
\footnote{See
  \texttt{http://swiftsc.gsfc.nasa.gov/docs/swift/proposals/cycle1\_gi.html}}
The satellite will be launched into a 22\degr-inclination low-Earth
orbit, and given the pointing restrictions of the BAT, XRT, and UVOT
instruments, there will be no ``continuous viewing zone.'' Rather, any
given position on the sky will be excluded from view for some or most
of each 95-minute orbit.  As a result -- and taking advantage of the
satellite's slewing capabilities -- each orbit will be divided into 4
to 6 distinct \mbox{10-min} to \mbox{20-min} pointings, each scheduled
during an appropriate visibility window.  In this manner the satellite
instruments will never suffer from Earth occultation, and a high duty
cycle for the mission can be achieved.  Protective measures for the
XRT and UVOT during passage through the South Atlantic Anomaly should
represent the only substantial recurring interruptions; even data that
are taken during satellite slews by the BAT and XRT instruments will
be amenable to later analysis.


\subsection{Relation to Primary Science}
\label{sub:swiftops:primary}

Through at least its first 1.5~years on-orbit, and quite possibly
beyond, \swift\ will be committed to the follow-up of every GRB it
detects, from the moment of trigger until the GRB has faded below
detectability to the XRT and UVOT.  This is expected to last one to
two weeks in the average case, although for rare, bright, low-redshift
events (e.g.\ GRB\,030329; \citealt{pfk+03,log+04}) it could last for
months.  The amount of time devoted to GRB follow-up, then, depends on
the expected trigger rate from the BAT.

Current estimates for this rate are in the range of 100 to 150 GRBs
per year.  Assuming \swift\ observation of five targets per orbit, and
ten days of follow-up observations per burst, this range of trigger
rates corresponds to a commitment of 55\% to 82\% of the total time
on-orbit.  The remainder of the on-orbit time will necessarily be
devoted to observations of non-afterglow targets.

The use of 10\% of the time on-orbit for the \swiftbscs\ thus seems
likely to be consistent with the primary GRB mission of \swift.  The
individual \mbox{500-s} pointings will be easy to fit into the typical
\mbox{20-min} visibility window for a single orbit; on average the
\swiftbscs\ will contribute one target per orbit to the \swift\
schedule.  When new GRBs are detected, the automated flight software
will override the preplanned observations and initiate a standard GRB
observations sequence, including follow-up observations in subsequent
orbits.  Since the \swiftbscs\ preplanned targets will always be of
lower priority than new GRBs and active afterglows, the execution of
the survey will at no point conflict with the GRB primary mission.


\subsection{Relation to Secondary Science}
\label{sub:swiftops:secondary}

Execution of the \swiftbscs\ is also consistent with the mission's
secondary science goals.  The most important non-GRB science goals of
\swift\ relate to the proposed all-sky surveys with the BAT
instrument.  Two survey components are planned: On a daily basis, the
BAT will be used to monitor the full sky down to source fluxes of
$\simgt$20\,mCrab (10--100\,keV); and over the course of the initial
two-year mission, a deep BAT all-sky survey will be conducted,
reaching 10--100\,keV source fluxes of $\simgt$1\,mCrab.  Both of
these surveys will be carried out by the BAT instrument team at
Goddard Space Flight Center and Los Alamos National Laboratory.  The
daily scans will allow the BAT to serve as a hard X-ray ``all-sky
monitor,'' alerting the community when a bright new source has
appeared or a familiar source has undergone a change of state; the
two-year all-sky survey will be $\simgt$10 times more sensitive than
the only previous such hard X-ray survey, by \textit{HEAO A-4} in
1977-79 \citep{lll+84}, and should detect hundreds of sources.

Since the BAT has a 1.4~sr field of view (half-coded), carrying out
the daily survey in a dedicated fashion would require $\approx$8
pointings, of 600\,s exposure each, to tile the accessible sky (given
the 45\degr\ exclusion cone around the Sun).  Some of this BAT survey
coverage will be provided by GRB follow-up observations, but at our
estimated trigger rates these provide only 3 to 5 active afterglows on
any given day.  The exact pointings selected to satisfy BAT coverage
of the remaining tiling positions will need to be chosen according to
some other criterion.

\begin{figure}[t]
\plotone{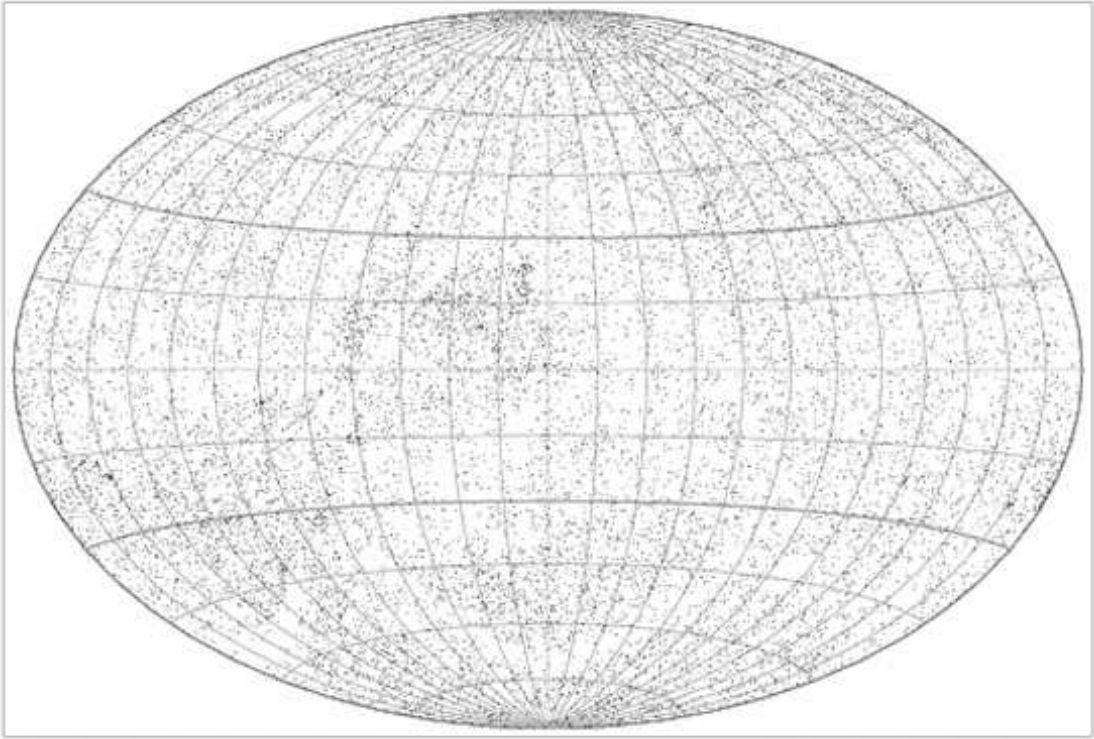}
\caption[]{\small%
Distribution of BSC sources in Galactic coordinates; Galactic
longitude $l=0\degr$ at the center and increases to the right.  The
catalog is complete over 92\% of the sky, with some gaps in coverage
due to the (ecliptic-oriented) scan pattern of the \rosat\ All-Sky
Survey (see \protect\citealt{vab+99}).  Lines of Galactic latitude
$|b|=30\degr$ are highlighted; these are the limits of the large-area
optical surveys for BSC counterparts that have been carried out to
date, which exclude the Galactic plane; see
\S\protect\ref{sub:promise:past} for details.
}
\label{fig:bsc-galactic}
\end{figure}

Since BSC sources provide close to uniform coverage of the full sky
(Fig.~\ref{fig:bsc-galactic}), targets at locations complementary to
the current GRB follow-up positions can be selected to complete the
tiling on each day -- the \mbox{500-s} exposure time of the
\swiftbscs\ is about the same as for the BAT pointings.  Observing BSC
sources at the appropriate tiling positions would account for 3 to 5
of the 17 BSC targets per day that will be observed for a survey using
10\% of the time on-orbit.


\subsection{Constraints}
\label{sub:swiftops:constraints}

There are several operational constraints on the \swift\ mission that
will impact its ability to carry out the \swiftbscs\ as we have
described it.

First, the current approved mission lifetime is two years, which is
not sufficient to complete the survey at a 10\% fraction of the time
on-orbit.  We consider the chances for a mission extension quite good,
however, given the unique capabilities of the satellite and the
revolution in GRB studies -- including the application of bright
afterglows to outstanding problems in astrophysics and cosmology --
that we expect it to bring about.  The satellite orbit itself is
expected to be stable for at least six years.

Next, changes in the UVOT filter involve moving parts with a finite
lifetime, so that it is desirable to minimize these filter changes.
Moreover, since the filters are arranged in a wheel with changes
happening in strict sequence, changes should preferentially be made in
this order.  During GRB follow-up observations, the six broadband
filters (and for bright bursts, the grisms) will be stepped through in
strict succession, with roughly equal exposure times per filter.  At
early times, filter changes will happen several times per visibility
window, while at late times, each visibility window will be devoted to
observations in a single filter.  

Since survey operations are of lower priority than afterglow
observations, we cannot expect to have free access to any particular
UVOT filter for any observation.  For this reason, we have avoided
assuming that the BSC sources will be observed uniformly in a single
filter, even though this would be the obvious preference for any
complete survey.  Moreover, in the case of the \swiftbscs, we have a
scientific preference for UV rather than optical observations: the
target BSC sources are bright and unextinguished, and likely to have
relatively bright UV emission; in addition, UV observations represent
a distinctive capability of \swift, while optical observations can be
made from the ground.  The ultimate allocation of filter changes for
the survey will thus represent a compromise between survey science and
satellite operations.

Next, given the satellite power constraints there is an approximate
500 degree per orbit slew limit.  In general the current operations
plan, which specifies 4 to 6 pointings per orbit to satisfy visibility
constraints, is not expected to tax this limit.  However, the
\mbox{500-s} observations of the \swiftbscs\ may be considered
slightly shorter than optimal, in the sense that they may not fill the
window of visibility for the target.  Given the ready availability of
BSC sources near most sky positions, it seems unlikely that the slew
constraint will present a problem for survey operations, and this can
be confirmed by simulations of the survey execution on orbit.

We note that with the \mbox{23.6-arcmin} field of view of the XRT it
will be possible in some cases to observe two BSC sources in a single
pointing.  This represents an opportunity to reduce the total time
request for the survey by approximately 10\%.

Finally, it is clear that the \swiftbscs\ cannot be carried out
without full support and close interaction with the \swift\ team.  In
a pragmatic sense, no such additional, large-scale project can be
contemplated without added moneys to support the activity.  We have
restricted ourselves to the scientific case for the \swiftbscs;
eventually, budgetary resources will also have to be secured.


\section{Conclusions}
\label{sec:conclusions}

We have investigated the prospects for a complete survey of the 18,811
sources of the \rosat\ All-Sky Survey Bright Source Catalog (BSC) with
the \swift\ gamma-ray burst satellite, the ``\swiftbscs,'' and find
them to be good.  With 10\% of the time on-orbit, over a three-year
period, each BSC source can be observed for 500~s with the satellite's
narrow-field X-ray (XRT) and UV/Optical (UVOT) telescopes.  The XRT
observation will provide an improved $<$5-arcsec position for every
BSC source that has not faded by more than a factor of 5 from its
\rosat-era flux level, and in addition, provide a 0.2--10 keV flux for
the source.  The localization will represent an improvement by a
factor of 10 to 30 in two-dimensional precision over the BSC positions
(Fig.~\ref{fig:bsc-gray}).  The simultaneous UVOT observation, taken
in one of six broadband filters in the 200~nm to 600~nm wavelength
range, will reach a limiting magnitude of $B=23.5$~mag or its rough
equivalent, yielding candidate UV/optical counterparts or deep limits
in every case.  

Given the brightness of the BSC sources, and the \lxlopt\ properties
of known source populations, the improved positions of the \swiftbscs\
can be expected to yield counterpart identifications in the great
majority of cases.  Indeed, \swiftbscs\ sources without plausible
optical (\usno\ or UVOT) counterparts will be immediate candidate
X-ray binaries or isolated neutron stars (INSs;
Fig~\ref{fig:bsc-hist}).

Deriving counterpart identifications for the great majority of BSC
sources will realize the promise of the BSC as a multiwavelength
catalog, and is the best means to enumerate the rare source
populations in the BSC and discover any new populations that may
exist.  Without improved positions these efforts will continue to be
frustrated by confusion problems -- with multiple \usno\ sources in
the typical 60\arcsec-diameter 2-sigma error circle, establishing a
unique identification is nontrivial \citep{rbp+00,rfb+03} -- and
affected by unknown levels of contamination -- most efforts have
associated each BSC source with the most promising candidate in its
error circle, without deriving a quantitative estimate of the
likelihood of association.  Each counterpart identification, then,
will enable flux estimates or upper limits across the electromagnetic
spectrum via current optical (DSS, DPOSS, SDSS), near-infrared
(2MASS), and radio (NVSS, FIRST) surveys.

Any interesting BSC source that can be identified in this manner will
be readily accessible to high signal-to-noise, high-resolution X-ray
spectroscopy by \xmmlong\ and \chandra.  In addition, those sources
with optical counterparts will for the most part be accessible to
optical-spectroscopic studies from modest-sized (\mbox{4-m} class)
ground-based facilities.  

We have investigated \swift\ mission constraints on the \swiftbscs\
and do not find them to be severe.  The BSC targets will be of lower
priority than new GRBs or active GRB afterglows, so that the survey
will not conflict with the primary \swift\ goal of GRB observations.
Moreover, the \swiftbscs\ observations can be scheduled so as to
advance the secondary science goals of a daily all-sky hard X-ray
(10--100 keV) survey to $\simgt$20\,mCrab, and a two-year all-sky
survey to $\simgt$1\,mCrab, executed with the wide-field Burst Alert
Telescope (BAT).  The 10\% commitment of satellite time corresponds to
roughly one BSC target per 95-min orbit.  Each such orbit will
typically involve observations at 4 to 6 distinct pointings, some 3 to
5 of which will be active GRB afterglows.

Execution of the \swiftbscs\ will surely represent an operational
challenge for the mission.  The progress of the survey will have to be
monitored on an orbit by orbit basis, with new targets chosen from the
unobserved portions of the BSC, and target lists updated as necessary
when new GRB triggers result in the automated rescheduling of
satellite observations.  Moreover, some level of global optimization
of target choices will be desired, to retain isotropy of the remaining
targets, as much as possible, until the very end of the survey.

However, the resulting capability for routine observations of large
numbers of distinct targets will prove valuable should the \swift\
team choose to pursue complete surveys of other target catalogs.  In
particular, the positions of sources identified in the two-year BAT
all-sky survey will only be known to few-arcmin precision, so that a
uniform campaign of follow-up observations with the XRT might be worth
considering.  Complete XRT+UVOT surveys of other target populations --
for example, targets from the forthcoming \textit{GALEX} all-sky
imaging survey -- might also be contemplated.  Execution of these or
similar projects will require a capability like that needed for the
\swiftbscs.

Ultimately, the choice to pursue the \swiftbscs\ is a choice to pursue
the science we have described in preference to other non-GRB
observations that \swift\ might carry out.  Beyond the first 1.5~years
of on-orbit operations, the \swift\ Guest Investigator program will
enter Cycle~2, and may solicit proposals from the community for
pointed XRT and UVOT observations of designated non-GRB targets.
Continued execution of the \swiftbscs\ into this era would imply that
less time would be available for these programs, for two or three
cycles, until the \swiftbscs\ is completed.

In a generic sense, the choice to make short observations of a large
number of targets, as we propose here, may be contrasted with the
alternative of making deeper exposures of some smaller set of targets,
for example, monitoring X-ray binaries, soft gamma-ray repeaters,
anomalous X-ray pulsars, or selected X-ray bright AGN at regular
intervals.  The variety of science which might be accomplished with
such campaigns is surely rich, and beyond our capability to summarize
or even hint at in this context.  However, we believe that the science
gains of a complete \swiftbscs\ are of sufficient value to warrant at
least a similar level of consideration.

Most importantly, the science of the \swiftbscs\ is science that only
\swift\ can achieve.  \swift's combination of multiwavelength
instrumentation, fast-slew capability, and flexibility in target
scheduling make it the only present or planned satellite mission
capable of executing a complete BSC survey without significant
operational overheads.  The \swiftbscs\ thus represents the best
opportunity for many years to come to expand the number of
securely-identified X-ray sources by an order of magnitude, from
$\sim$2000 to $\sim$18,000.  Declining to take advantage of this
opportunity will restrict the observations of the current and
next-generation X-ray missions to more or less the known population of
X-ray sources that exists today.  On the other hand, prompt execution
of the \swiftbscs\ will allow immediate follow-up studies with \xmm\
and \chandra\ of the most intriguing new members of all X-ray source
populations, from stars to compact objects to AGN.  This will renew
the promise of these observatories, for all areas of X-ray astronomy,
as they enter their extended mission phase.


\section*{Acknowledgments.}
The author would like to express his appreciation to Neil Gehrels,
John Nousek, David Burrows, and the \swift\ mission team for the
advance planning and extensive consideration of mission capabilities
that have made this proposal possible.  In addition, he thanks Bob
Rutledge, Dale Frail, and Shri Kulkarni for helpful discussions.


\clearpage


\begin{table}
\begin{centering}
\begin{tabular}{lll}
Property               &    PSPC   & Swift      \\ \hline
Effective Area (\cmsq) &     150   &   135      \\
Energy range (keV)     & 0.1--2.4 & 0.2--10.0   \\
PSF FWHM (arcsec)      &   25.0    &   18.0     \\
$E/\Delta E$           &   2.5     & $\simgt$10 \\ \hline
\end{tabular}
\caption{%
Basic properties of the \rosat\ PSPC and \swift\ XRT.\label{tab:props}}
\end{centering}
\end{table}

		
\begin{thebibliography}{22}
\expandafter\ifx\csname natexlab\endcsname\relax\def\natexlab#1{#1}\fi

\bibitem[{{Anderson} {et~al.}(2003){Anderson}, {Voges}, {Margon}, {Tr{\"
  u}mper}, {Ag{\" u}eros}, {Boller}, {Collinge}, {Homer}, {Stinson}, {Strauss},
  {Annis}, {G{\' o}mez}, {Hall}, {Nichol}, {Richards}, {Schneider}, {Vanden
  Berk}, {Fan}, {Ivezi{\' c}}, {Munn}, {Newberg}, {Richmond}, {Weinberg},
  {Yanny}, {Bahcall}, {Brinkmann}, {Fukugita}, \& {York}}]{avm+03}
{Anderson}, S.~F. {et al.}\  2003, AJ, 126, 2209

\bibitem[{{Bade} {et~al.}(1998){Bade}, {Engels}, {Voges}, {Beckmann}, {Boller},
  {Cordis}, {Dahlem}, {Englhauser}, {Molthagen}, {Nass}, {Studt}, \&
  {Reimers}}]{bev+98}
{Bade}, N. {et al.}\  1998, 127, 145

\bibitem[{{Barthelmy}(2000)}]{barthelmy00}
{Barthelmy}, S.~D. 2000, in Proc. SPIE Vol. 4140, p. 50-63, X-Ray and Gamma-Ray
  Instrumentation for Astronomy XI, Kathryn A. Flanagan; Oswald H. Siegmund;
  Eds., 50--63

\bibitem[{{Burrows} {et~al.}(2003){Burrows}, {Hill}, {Nousek}, {Wells},
  {Short}, {Ambrosi}, {Chincarini}, {Citterio}, \& {Tagliaferri}}]{bhn+03}
{Burrows}, D.~N. {et al.}\  2003, in X-Ray and Gamma-Ray Telescopes and
  Instruments for Astronomy. Edited by Joachim E. Truemper, Harvey D.
  Tananbaum. Proceedings of the SPIE, Volume 4851, pp. 1320-1325 (2003).,
  1320--1325

\bibitem[{{Fischer} {et~al.}(1998){Fischer}, {Hasinger}, {Schwope}, {Brunner},
  {Boller}, {Trumper}, {Voges}, \& {Neizvestny}}]{fhs+98}
{Fischer}, J.-U., {Hasinger}, G., {Schwope}, A.~D., {Brunner}, H., {Boller},
  T., {Trumper}, J., {Voges}, W., \& {Neizvestny}, S. 1998, Astronomische
  Nachrichten, 319, 347

\bibitem[{{Kaspi} {et~al.}(2004){Kaspi}, {Roberts}, \& {Harding}}]{krh04}
{Kaspi}, V.~M., {Roberts}, M.~S.~E., \& {Harding}, A.~K. 2004, ArXiv
  Astrophysics e-prints, astro-ph/0402136

\bibitem[{{Levine} {et~al.}(1984){Levine}, {Lang}, {Lewin}, {Primini},
  {Dobson}, {Doty}, {Hoffman}, {Howe}, {Scheepmaker}, {Wheaton}, {Matteson},
  {Baity}, {Gruber}, {Knight}, {Nolan}, {Pelling}, {Rothschild}, \&
  {Peterson}}]{lll+84}
{Levine}, A.~M. {et al.}\  1984, \apjs, 54, 581

\bibitem[{{Lipkin} {et~al.}(2003){Lipkin}, {Ofek}, {Gal-Yam}, {Leibowitz},
  {Poznanski}, {Kaspi}, {Polishook}, {Kulkarni}, {Fox}, {Berger}, {Mirabal},
  {Halpern}, {Bureau}, {Fathi}, {Price}, {Peterson}, {Frebel}, {Schmidt},
  {Orosz}, {Fitzgerald}, {Bloom}, {van Dokkum}, {Bailyn}, {Buxton}, \&
  {Barsony}}]{log+04}
{Lipkin}, Y.~M. {et al.}\  2003, astro-ph/0312594

\bibitem[{{Mason} {et~al.}(2001){Mason}, {Breeveld}, {Much}, {Carter},
  {Cordova}, {Cropper}, {Fordham}, {Huckle}, {Ho}, {Kawakami}, {Kennea},
  {Kennedy}, {Mittaz}, {Pandel}, {Priedhorsky}, {Sasseen}, {Shirey}, {Smith},
  \& {Vreux}}]{mbm+01}
{Mason}, K.~O. {et al.}\  2001, \aap, 365, L36

\bibitem[{{Motch} {et~al.}(1999){Motch}, {Haberl}, {Zickgraf}, {Hasinger}, \&
  {Schwope}}]{mhz+99}
{Motch}, C., {Haberl}, F., {Zickgraf}, F.-J., {Hasinger}, G., \& {Schwope},
  A.~D. 1999, A\&A, 351, 177

\bibitem[{{Price} {et~al.}(2003){Price}, {Fox}, {Kulkarni}, {Peterson},
  {Schmidt}, {Soderberg}, {Yost}, {Berger}, {Djorgovski}, {Frail}, {Harrison},
  {Sari}, {Blain}, \& {Chapman}}]{pfk+03}
{Price}, P.~A. {et al.}\  2003, Nature, 423, 844

\bibitem[{{Roming} {et~al.}(2000){Roming}, {Townsley}, {Nousek}, {Altimore},
  {Case}, {Hunsberger}, {Koch}, {Mason}, {Carter}, {Cropper}, {Hancock},
  {Huckle}, {Kennedy}, {McLelland}, {Smith}, {Killough}, \& {Ho}}]{rtn+00}
{Roming}, P.~W. {et al.}\  2000, in Proc. SPIE Vol. 4140, p. 76-86, X-Ray and
  Gamma-Ray Instrumentation for Astronomy XI, Kathryn A. Flanagan; Oswald H.
  Siegmund; Eds., 76--86

\bibitem[{{Rutledge} {et~al.}(2000){Rutledge}, {Brunner}, {Prince}, \&
  {Lonsdale}}]{rbp+00}
{Rutledge}, R.~E., {Brunner}, R.~J., {Prince}, T.~A., \& {Lonsdale}, C. 2000,
  \apjs, 131, 335

\bibitem[{{Rutledge} {et~al.}(2003){Rutledge}, {Fox}, {Bogosavljevic}, \&
  {Mahabal}}]{rfb+03}
{Rutledge}, R.~E., {Fox}, D.~W., {Bogosavljevic}, M., \& {Mahabal}, A. 2003,
  ApJ, 598, 458

\bibitem[{{Schwope} {et~al.}(2000){Schwope}, {Hasinger}, {Lehmann}, {Schwarz},
  {Brunner}, {Neizvestny}, {Ugryumov}, {Balega}, {Tr{\" u}mper}, \&
  {Voges}}]{shl+00}
{Schwope}, A. {et al.}\  2000, Astronomische Nachrichten, 321, 1

\bibitem[{{Schwope} {et~al.}(1999){Schwope}, {Hasinger}, {Schwarz}, {Haberl},
  \& {Schmidt}}]{shs+99}
{Schwope}, A.~D., {Hasinger}, G., {Schwarz}, R., {Haberl}, F., \& {Schmidt}, M.
  1999, A\&A, 341, L51

\bibitem[{{Treves} {et~al.}(2000){Treves}, {Turolla}, {Zane}, \&
  {Colpi}}]{ttz+00}
{Treves}, A., {Turolla}, R., {Zane}, S., \& {Colpi}, M. 2000, PASP, 112, 297

\bibitem[{{Voges} {et~al.}(1999){Voges}, {Aschenbach}, {Boller}, {Br{\"
  a}uninger}, {Briel}, {Burkert}, {Dennerl}, {Englhauser}, {Gruber}, {Haberl},
  {Hartner}, {Hasinger}, {K{\" u}rster}, {Pfeffermann}, {Pietsch}, {Predehl},
  {Rosso}, {Schmitt}, {Tr{\" u}mper}, \& {Zimmermann}}]{vab+99}
{Voges}, W. {et al.}\  1999, A\&A, 349, 389

\bibitem[{{Voges} {et~al.}(2000){Voges}, {Aschenbach}, {Boller}, {Brauninger},
  {Briel}, {Burkert}, {Dennerl}, {Englhauser}, {Gruber}, {Haberl}, {Hartner},
  {Hasinger}, {Pfeffermann}, {Pietsch}, {Predehl}, {Schmitt}, {Trumper}, \&
  {Zimmermann}}]{vab+00}
{Voges}, W. {et al.}\  2000, in International Astronomical Union Circular, 1--+

\bibitem[{{Walter} \& {Matthews}(1997)}]{wm97}
{Walter}, F.~M. \& {Matthews}, L.~D. 1997, Nature, 389, 358

\bibitem[{{York} {et~al.}(2000){York}, {Adelman}, {Anderson}, {Anderson},
  {Annis}, {Bahcall}, {Bakken}, {Barkhouser}, {Bastian}, {Berman}, {Boroski},
  {Bracker}, {Briegel}, {Briggs}, {Brinkmann}, {Brunner}, {Burles}, {Carey},
  {Carr}, {Castander}, {Chen}, {Colestock}, {Connolly}, {Crocker}, {Csabai},
  {Czarapata}, {Davis}, {Doi}, {Dombeck}, {Eisenstein}, {Ellman}, {Elms},
  {Evans}, {Fan}, {Federwitz}, {Fiscelli}, {Friedman}, {Frieman}, {Fukugita},
  {Gillespie}, {Gunn}, {Gurbani}, {de Haas}, {Haldeman}, {Harris}, {Hayes},
  {Heckman}, {Hennessy}, {Hindsley}, {Holm}, {Holmgren}, {Huang}, {Hull},
  {Husby}, {Ichikawa}, {Ichikawa}, {Ivezi{\' c}}, {Kent}, {Kim}, {Kinney},
  {Klaene}, {Kleinman}, {Kleinman}, {Knapp}, {Korienek}, {Kron}, {Kunszt},
  {Lamb}, {Lee}, {Leger}, {Limmongkol}, {Lindenmeyer}, {Long}, {Loomis},
  {Loveday}, {Lucinio}, {Lupton}, {MacKinnon}, {Mannery}, {Mantsch}, {Margon},
  {McGehee}, {McKay}, {Meiksin}, {Merelli}, {Monet}, {Munn}, {Narayanan},
  {Nash}, {Neilsen}, {Neswold}, {Newberg}, {Nichol}, {Nicinski}, {Nonino},
  {Okada}, {Okamura}, {Ostriker}, {Owen}, {Pauls}, {Peoples}, {Peterson},
  {Petravick}, {Pier}, {Pope}, {Pordes}, {Prosapio}, {Rechenmacher}, {Quinn},
  {Richards}, {Richmond}, {Rivetta}, {Rockosi}, {Ruthmansdorfer}, {Sandford},
  {Schlegel}, {Schneider}, {Sekiguchi}, {Sergey}, {Shimasaku}, {Siegmund},
  {Smee}, {Smith}, {Snedden}, {Stone}, {Stoughton}, {Strauss}, {Stubbs},
  {SubbaRao}, {Szalay}, {Szapudi}, {Szokoly}, {Thakar}, {Tremonti}, {Tucker},
  {Uomoto}, {Vanden Berk}, {Vogeley}, {Waddell}, {Wang}, {Watanabe},
  {Weinberg}, {Yanny}, \& {Yasuda}}]{yaa+00}
{York}, D.~G. {et al.}\  2000, \aj, 120, 1579

\bibitem[{{Zickgraf} {et~al.}(2003){Zickgraf}, {Engels}, {Hagen}, {Reimers}, \&
  {Voges}}]{zeh+03}
{Zickgraf}, F.-J., {Engels}, D., {Hagen}, H.-J., {Reimers}, D., \& {Voges}, W.
  2003, A\&A, 406, 535

\end{thebibliography}
\end{document}